\documentclass[prd,reprint,showpacs,showkeys]{revtex4-1}
\usepackage{amsfonts}
\usepackage{amssymb}
\usepackage{amsmath}
\usepackage{graphicx}
\usepackage[font={footnotesize,it}]{caption}

\begin{document}

\title{Black holes from multiplets of scalar fields in 2+1- and
3+1-dimensions}
\author{S. Habib Mazharimousavi}
\email{habib.mazhari@emu.edu.tr}
\author{M. Halilsoy}
\email{mustafa.halilsoy@emu.edu.tr}
\affiliation{Department of Physics, Eastern Mediterranean University, Gazima\~{g}usa,
Turkey. }
\date{\today }

\begin{abstract}
We obtain classes of black hole solutions constructed from multiplets of
scalar fields in 2+1 / 3+1 dimensions. The multi-component scalars don't
undergo a symmetry breaking so that only the isotropic modulus is effective.
The Lagrangian is supplemented by a self-interacting potential which plays
significant role in obtaining the exact solutions. In 2+1 / 3+1 dimensions
doublet / triplet of scalars is effective which enriches the available black
hole spacetimes and creates useful Liouville weighted field theoretic models.
\end{abstract}

\maketitle

\section{Introduction}

The absence of gravitational degrees in lower dimensions stipulates addition
of physical sources in order to make strong attraction centers and black
holes. The prototype example in this regard was provided by the Ba\~{n}%
ados-Teitelboim-Zanelli (BTZ) black hole in $2+1-$dimensions which was
sourced by a cosmological constant \cite{1,2}. Addition of different sources
to make alternative black holes to the BTZ has always been challenging \cite%
{3,4,5}. From this token recently we considered a doublet of scalar fields
constrained to lie on the unit sphere as source in $2+1-$dimensions \cite{6}%
. The uniqueness condition of the scalar fields under rotation imposes an
integer parameter to play role in the metric. Two distinct classes of
solutions emerged: a black hole with integer valued negative Hawking
temperature and a non-black hole metric with interesting topological
properties. Although extension of similar properties to $3+1-$dimensional
metrics remains to be seen it is of utmost importance in connection with the
belief that spacetime may be 'digital'. At the quantum (Planck) level the
idea is not new but at the classical, large scale it needs concrete proof to
incorporate topological numbers. Beside black holes wormholes also can be
considered within the similar context. In a recent work we have shown for
instance that a wormhole solution can also be obtained by employing a
scalar-doublet of fields as source in $2+1-$dimensions \cite{7}. Let's add
that multiple field scalar-tensor has been studied before \cite%
{8,9,10,11,12,13,14,15,16,17,18,19}. Triplet scalar field in the context of
global monopole has been studied extensively in literature \cite%
{20,21,22,23,24,25,26,27,28,29}. One has to keep in order that the single
scalar field coupled with gravity has been studied more rigorously. As our
concentration is on $2+1-$ and $3+1-$ dimensions, we only refer to \cite%
{30,31} in $2+1-$ and \cite{32,33,34,35,36,37,38,39,40} in $3+1-$dimensions.

In the present study we choose firstly our source again as a doublet of
scalar fields, namely $\phi ^{1}\left( r,\theta \right) =\phi \left(
r\right) \cos \theta $ and $\phi ^{2}\left( r,\theta \right) =\phi \left(
r\right) \sin \theta ,$ with the modulus $\phi \left( r\right) .$ Our metric
is circularly symmetric so that the angular dependence washes out leaving
only the radial dependent function $\phi \left( r\right) .$ In addition to
the kinetic term of the scalar field we choose a suitable potential term
such that our system will admit a black hole solution with interesting
properties. The chosen potential with $V\left( \phi \right) $ is the product
of a polynomial expression with a Liouville term. The number of parameters
initially is four but with the solution the number reduces to two. The
potential admits a local minimum apt to define a vacuum in the assumed field
theory model. Particle states can be constructed in the potential well in
analogy with the energy levels of atoms. The potential has the constant term
with $\phi \left( r\right) =0$, which leads to the well-known BTZ black
hole. Our solution can be interpreted as a new black hole solution
constructed from a simple doublet of scalar fields. Such black holes emerge
with distinct properties when compared with a singlet scalar field black
holes. Secondly, we undertake the similar task to construct black holes in $%
3+1-$dimensions whose source consists of a triplet of scalar fields. The
solution in the limit of zero scalar fields naturally reduces to the
Schwarzschild-de Sitter spacetime.

Organization of the paper is as follows. Section II and III are devoted to
the $2+1-$dimensional field theoretic black hole solutions. Parallel
considerations for $3+1-$dimensions will be analyzed in Sections IV and V.
Our brief conclusion in Section VI completes the paper.

\section{$2+1-$dimensional field equations}

Our action in $2+1-$dimensional gravity minimally coupled to a doublet of
scalar field and without cosmological constant is given by ($16\pi G=c=1$)%
\begin{equation}
S=\int d^{3}x\sqrt{-g}\left( R+\mathcal{L}_{field}\right)
\end{equation}%
in which 
\begin{equation}
\mathcal{L}_{field}=-\frac{1}{2}\left( \nabla \phi ^{a}\right) ^{2}-V\left(
\phi \right) .
\end{equation}%
Here 
\begin{equation}
\left( 
\begin{array}{c}
\phi ^{1} \\ 
\phi ^{2}%
\end{array}%
\right) =\phi \left( r\right) \left( 
\begin{array}{c}
\cos \theta \\ 
\sin \theta%
\end{array}%
\right)
\end{equation}%
is the doublet of scalar fields with modulus%
\begin{equation}
\phi =\phi \left( r\right) =\pm \sqrt{\phi ^{a}\phi ^{a}}
\end{equation}%
and 
\begin{equation}
V\left( \phi \right) =V_{0}\left( 1+\xi _{1}\phi +\xi _{2}\phi ^{2}\right)
e^{-\alpha \phi }
\end{equation}%
is our potential ansatz with real parameters $V_{0},$ $\xi _{1},$ $\xi _{2},$
and $\alpha \in 
\mathbb{R}
-\left\{ \pm \sqrt{2}\right\} .$ Let's add that with specific form of
potential (5), $\alpha =\pm \sqrt{2}$ does not admit a solution to our field
equations therefore from the outset we exclude it. The trivial solution
comes with $\phi =0$ case where we have $V\left( \phi \right) =V_{0}$ which
can be considered as a cosmological constant to yield the BTZ solution.

The circularly symmetric line element is chosen to be 
\begin{equation}
ds^{2}=-A\left( r\right) dt^{2}+\frac{1}{A(r)}dr^{2}+H\left( r\right)
^{2}d\theta ^{2}
\end{equation}%
in which $A\left( r\right) $ and $H\left( r\right) $ are two functions only
of $r$. The field Lagrangian density may be cast into the following explicit
form%
\begin{equation}
\mathcal{L}_{field}=-\frac{A}{2}\phi ^{\prime 2}-\frac{1}{2H^{2}}\phi
^{2}-V\left( \phi \right) ,
\end{equation}%
whose variation with respect to $\phi $ yields the corresponding field
equation%
\begin{equation}
\phi ^{\prime \prime }+\frac{\left( AH\right) ^{\prime }}{AH}\phi ^{\prime }-%
\frac{\phi }{AH^{2}}-\frac{V^{\prime }\left( \phi \right) }{A}=0.
\end{equation}%
We note that a prime stands for a derivative with respect to the argument of
the function. Furthermore, variation of the action with respect to $g^{\mu
\nu }$ gives the Einstein's field equations%
\begin{equation}
G_{\mu }^{\nu }=T_{\mu }^{\nu }
\end{equation}%
in which the energy momentum tensor $T_{\mu }^{\nu }$ is defined as 
\begin{equation}
T_{\mu }^{\nu }=\frac{1}{2}\left( \partial _{\mu }\phi ^{a}\partial ^{\nu
}\phi ^{a}-\frac{1}{2}\left( \nabla \phi ^{a}\right) ^{2}\delta _{\mu }^{\nu
}\right) -\frac{1}{2}V\left( \phi \right) \delta _{\mu }^{\nu }.
\end{equation}%
One may find the nonzero components of $T_{\mu }^{\nu }$ given by 
\begin{equation}
T_{t}^{t}=-\frac{1}{4}\left( A\phi ^{\prime 2}+\frac{\phi ^{2}}{H^{2}}%
\right) -\frac{1}{2}V\left( \phi \right) ,
\end{equation}%
\begin{equation}
T_{r}^{r}=\frac{1}{4}\left( A\phi ^{\prime 2}-\frac{\phi ^{2}}{H^{2}}\right)
-\frac{1}{2}V\left( \phi \right)
\end{equation}%
and%
\begin{equation}
T_{\theta }^{\theta }=-\frac{1}{4}\left( A\phi ^{\prime 2}-\frac{\phi ^{2}}{%
H^{2}}\right) -\frac{1}{2}V\left( \phi \right) .
\end{equation}%
Finally the explicit form of the Einstein's equations are given by%
\begin{equation}
\frac{2H^{\prime \prime }A+A^{\prime }H^{\prime }}{H}+\frac{1}{2}\left(
A\phi ^{\prime 2}+\frac{\phi ^{2}}{H^{2}}\right) +V=0,
\end{equation}%
\begin{equation}
\frac{A^{\prime }H^{\prime }}{H}-\frac{1}{2}\left( A\phi ^{\prime 2}-\frac{%
\phi ^{2}}{H^{2}}\right) +V=0
\end{equation}%
and%
\begin{equation}
A^{\prime \prime }+\frac{1}{2}\left( A\phi ^{\prime 2}-\frac{\phi ^{2}}{H^{2}%
}\right) +V=0
\end{equation}%
which together with (8) must be solved. In short we seek for a set of
functions including $A,$ $H$ and $\phi $ which satisfy the four coupled
differential equations given by (8) and (14-16).

\section{Solution to the field equations in $2+1-$dimensions}

To solve the field equations, first we combine the $tt$ and $rr$ components
of the Einstein's equations to find%
\begin{equation}
2H^{\prime \prime }+\phi ^{\prime 2}H=0.
\end{equation}%
Next we consider an ansatz for $H$ given by%
\begin{equation}
H=H_{0}e^{\frac{\alpha }{2}\phi }
\end{equation}%
with $H_{0}$ and $\alpha $ two parameters to be found. The latter choice and
(17) yield a solution for $\phi $ as%
\begin{equation}
\phi =\frac{2\alpha }{\alpha ^{2}+2}\ln \left( C_{1}r+C_{2}\right)
\end{equation}%
in which $C_{1}$ and $C_{2}$ are two integration constants. We note that by
introducing $\bar{r}=C_{1}r+C_{2}$ one finds $d\bar{r}=C_{1}dr$ and $\phi $
rescaled. This however, does not bring new contribution to the problem.
Therefore without loss of generality \cite{41} we set $C_{1}=1$ and $C_{2}=0$
which yields%
\begin{equation}
\phi =\frac{2\alpha }{\alpha ^{2}+2}\ln r.
\end{equation}%
Plugging $\phi $ and $H$ into $tt$ equation we find the solution for $%
A\left( r\right) $ which must satisfy the other field equations too. Doing
this, however, imposes that 
\begin{equation}
\xi _{1}=\frac{\alpha ^{2}-2}{\alpha },
\end{equation}%
\begin{equation}
\xi _{2}=\frac{\left( \alpha ^{2}-2\right) ^{2}}{\alpha ^{4}},
\end{equation}%
\begin{equation}
H_{0}^{2}=\frac{\alpha ^{4}}{V_{0}\left( \alpha ^{2}-2\right) ^{3}}.
\end{equation}%
From this point on we shall make the choice $H_{0}=1$ so that the constant $%
V_{0}$ will be expressed in terms of $\alpha $, namely 
\begin{equation}
V_{0}=\frac{\alpha ^{4}}{\left( \alpha ^{2}-2\right) ^{3}}.
\end{equation}%
Therefore, a complete set of solutions to the field equations are given by%
\begin{equation}
H\left( r\right) =r^{\frac{\alpha ^{2}}{\alpha ^{2}+2}},
\end{equation}%
\begin{equation}
\phi =\frac{2\alpha }{\alpha ^{2}+2}\ln r,
\end{equation}%
\begin{equation}
V\left( \phi \right) =\frac{\alpha ^{4}}{\left( \alpha ^{2}-2\right) ^{3}}%
\left( 1+\frac{\alpha ^{2}-2}{\alpha }\phi +\frac{\left( \alpha
^{2}-2\right) ^{2}}{\alpha ^{4}}\phi ^{2}\right) e^{-\alpha \phi }
\end{equation}%
with the metric function%
\begin{multline}
A\left( r\right) =C_{0}r^{\frac{2}{\alpha ^{2}+2}} \\
-\frac{\alpha ^{2}r^{\frac{4}{\alpha ^{2}+2}}}{\alpha ^{2}-2}\left( \ln
^{2}r-\frac{\left( \alpha ^{2}+2\right) \left( \alpha ^{2}-3\right) \ln r}{%
\left( \alpha ^{2}-2\right) }\right. \\
+\left. \frac{\left( \alpha ^{2}+2\right) ^{2}\left( \alpha ^{4}-5\alpha
^{2}+7\right) }{2\left( \alpha ^{2}-2\right) ^{2}}\right) ,
\end{multline}%
in which $C_{0}$ is an integration constant. The Kretschmann scalar of the
solution can be written as%
\begin{multline}
K=\omega _{1}r^{-\frac{4\alpha ^{2}}{\alpha ^{2}+2}}\ln ^{4}r+\omega _{2}r^{-%
\frac{4\alpha ^{2}}{\alpha ^{2}+2}}\ln ^{3}r+ \\
\left( \omega _{3}r^{-\frac{4\alpha ^{2}+2}{\alpha ^{2}+2}}+\omega _{4}r^{-%
\frac{4\alpha ^{2}}{\alpha ^{2}+2}}\right) \ln ^{2}r+ \\
\left( \omega _{5}r^{-\frac{4\alpha ^{2}+2}{\alpha ^{2}+2}}+\omega _{6}r^{-%
\frac{4\alpha ^{2}}{\alpha ^{2}+2}}\right) \ln r+ \\
\omega _{7}r^{-\frac{4\alpha ^{2}+2}{\alpha ^{2}+2}}+\omega _{8}r^{-\frac{%
4\alpha ^{2}+4}{\alpha ^{2}+2}}+\omega _{9}r^{-\frac{4\alpha ^{2}}{\alpha
^{2}+2}}
\end{multline}%
in which $\omega _{i}$ are regular functions of $\alpha $ and $C_{0}.$ As we
observe here the only singular point is the origin.

The solution for the metric function admits non-asymptotically flat black
hole solutions. A transformation of the form $r=\rho ^{\frac{\alpha ^{2}+2}{%
\alpha ^{2}}}$ makes the line element (6) to be of the form%
\begin{equation}
ds^{2}=-A\left( \rho \right) dt^{2}+\frac{\left( 1+\frac{2}{\alpha ^{2}}%
\right) ^{2}\rho ^{\frac{4}{\alpha ^{2}}}}{A(\rho )}d\rho ^{2}+\rho
^{2}d\theta ^{2}
\end{equation}%
in which 
\begin{multline}
A\left( \rho \right) =C_{0}^{\frac{2}{\alpha ^{2}}}\rho -\frac{\left( \alpha
^{2}+2\right) ^{2}}{\alpha ^{2}\left( \alpha ^{2}-2\right) }\rho ^{\frac{4}{%
\alpha ^{2}}} \\
\times \left( \ln ^{2}\rho -\frac{\alpha ^{2}\left( \alpha ^{2}-3\right) }{%
\left( \alpha ^{2}-2\right) }\ln \rho \right. \\
+\left. \frac{\alpha ^{4}\left( \alpha ^{4}-5\alpha ^{2}+7\right) }{2\left(
\alpha ^{2}-2\right) ^{2}}\right) .
\end{multline}%
Moreover, the scalar field $\phi $ becomes simply%
\begin{equation}
\phi =\frac{2}{\alpha }\ln \rho .
\end{equation}%
It is needless to state that the parameter $\alpha $ ($0<\alpha <\infty
,\alpha ^{2}\neq 2$) represents the scalar hair of the black hole.

To complete this section we find the quasi local mass of the central black
hole by applying the Brown and York (BY) formalism \cite{42,43}. This
technique is used for non-asymptotically flat spherically symmetric black
hole solution where an ADM mass may not be defined. According to \cite{42,43}
for a spherically symmetric N-dimensional spacetime%
\begin{equation}
ds^{2}=-F\left( \rho \right) ^{2}dt^{2}+\frac{d\rho ^{2}}{G\left( \rho
\right) ^{2}}+\rho ^{2}d\Omega _{N-2}^{2}
\end{equation}%
the quasilocal mass is defined to be%
\begin{equation}
M_{QL}=\lim_{\rho _{B}\rightarrow \infty }\frac{N-2}{2}\rho
_{B}^{N-3}F\left( \rho _{B}\right) \left( G_{ref}\left( \rho _{B}\right)
-G\left( \rho _{B}\right) \right)
\end{equation}%
in which $G_{ref}\left( \rho _{B}\right) $ is an arbitrary non-negative
reference function and $\rho _{B}$ is the radius of the spacelike
hypersurface boundary which is going to be infinite. In our case ($N=3$) we
have 
\begin{equation}
F\left( \rho _{B}\right) ^{2}=A\left( \rho _{B}\right)
\end{equation}%
\begin{equation}
G\left( \rho _{B}\right) ^{2}=\frac{A\left( \rho _{B}\right) \rho _{B}^{-%
\frac{4}{\alpha ^{2}}}}{\left( 1+\frac{2}{\alpha ^{2}}\right) ^{2}}
\end{equation}%
and by assuming that $A\left( \rho _{B}\right) $ diverges faster than $\rho
_{B}^{\frac{2}{\alpha ^{2}}}$ one finds 
\begin{equation}
M_{QL}=-\frac{C_{0}}{4\left( 1+\frac{2}{\alpha ^{2}}\right) }.
\end{equation}%
%
%
%
%
%
%
%
%
%
%
%
%
\begin{figure}[h]
\includegraphics[width=65mm,scale=0.7]{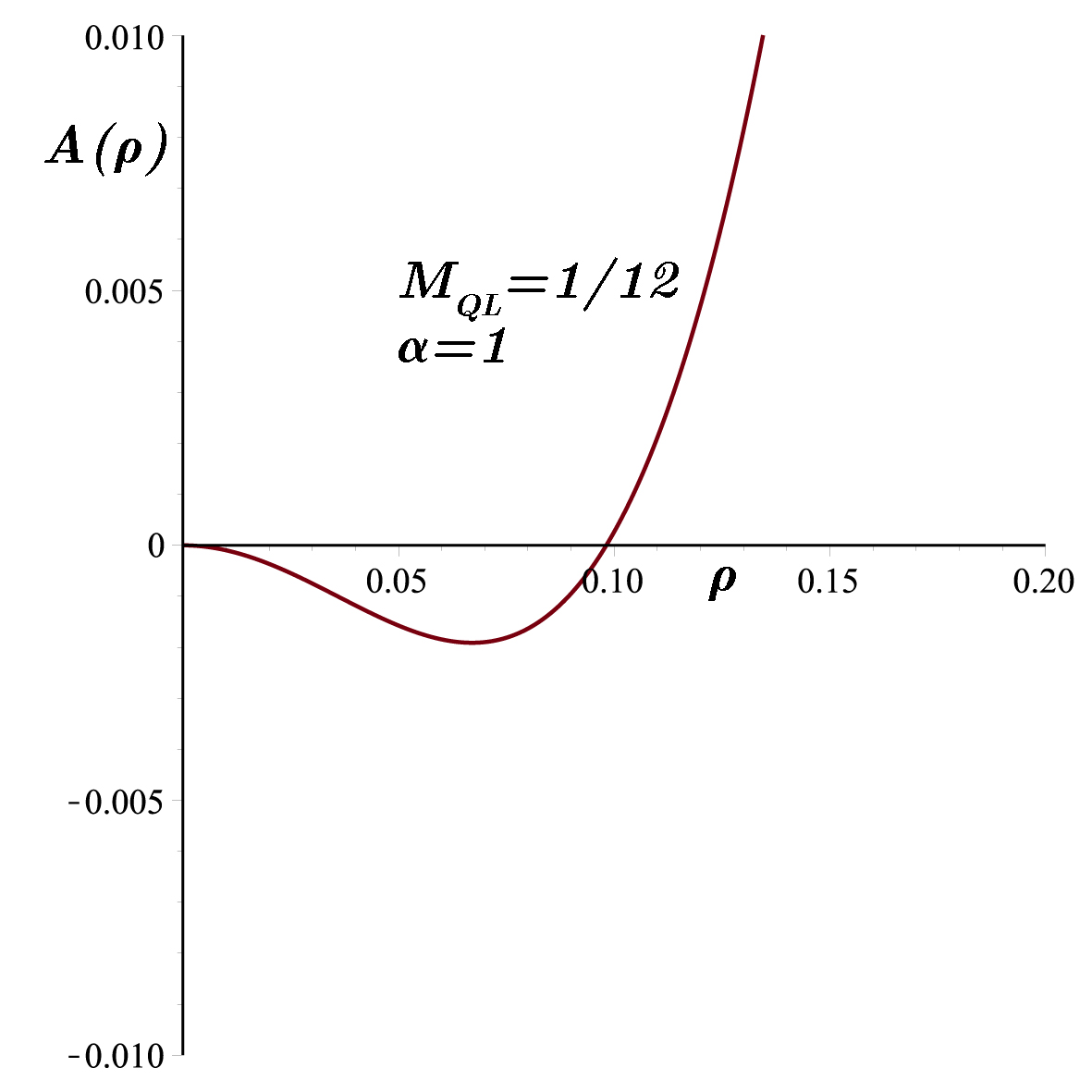}
\caption{Metric $A\left( \protect\rho \right) $ function versus $\protect%
\rho $ for $M_{QL}=\frac{1}{12}$ and $\protect\alpha =1$. The black hole is
not asymptotically flat (Eq. 43).}
\label{fig: 1}
\end{figure}
\begin{figure}[h]
\includegraphics[width=65mm,scale=0.7]{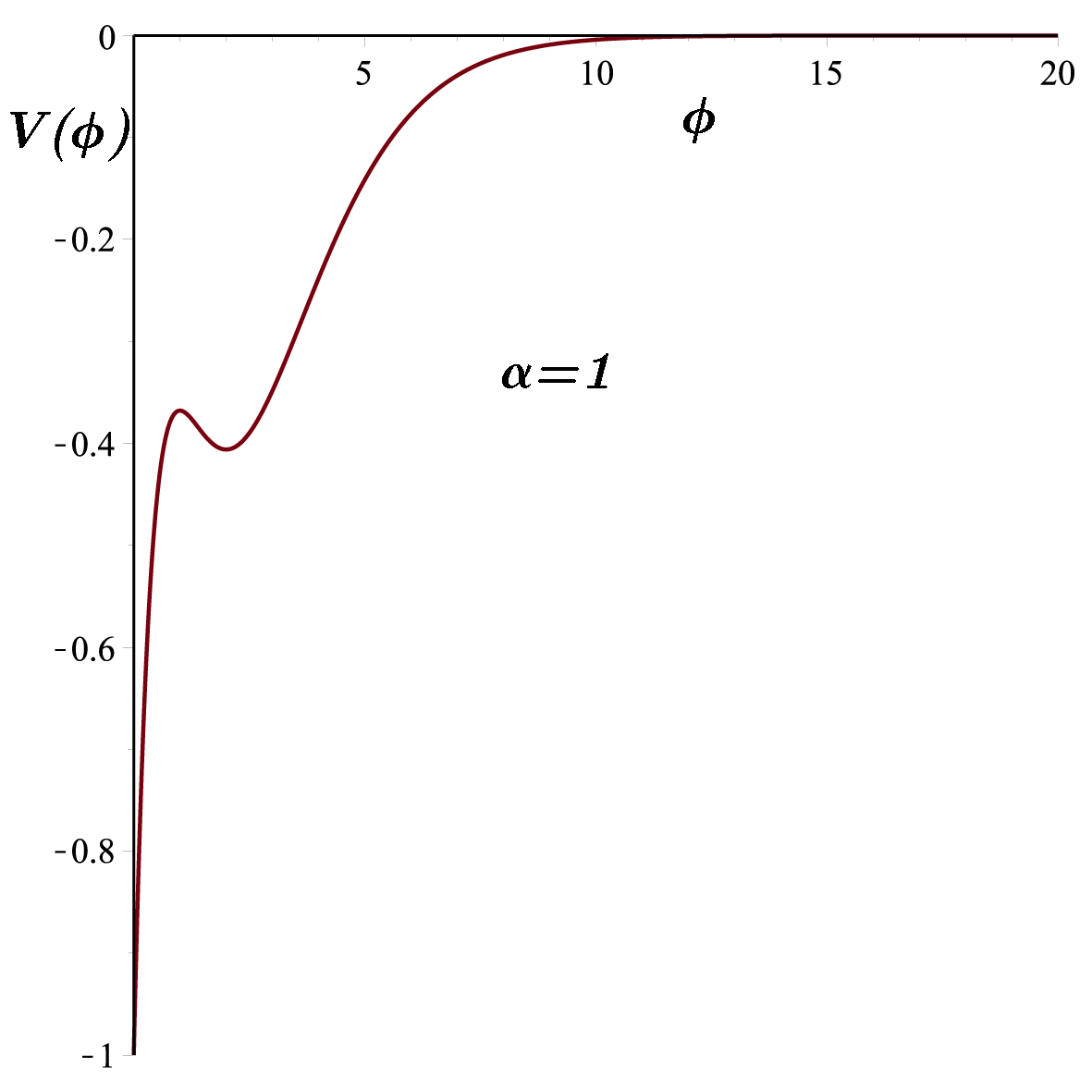}
\caption{The self-coupling potential $V\left( \protect\phi \right) $ versus $%
\protect\phi $ for $M_{QL}=\frac{1}{12}$ and $\protect\alpha =1.$ The
minimum of the potential is the stability point of the scalar field (Eq.
27). }
\label{fig: 2}
\end{figure}

\begin{figure}[h]
\includegraphics[width=65mm,scale=0.7]{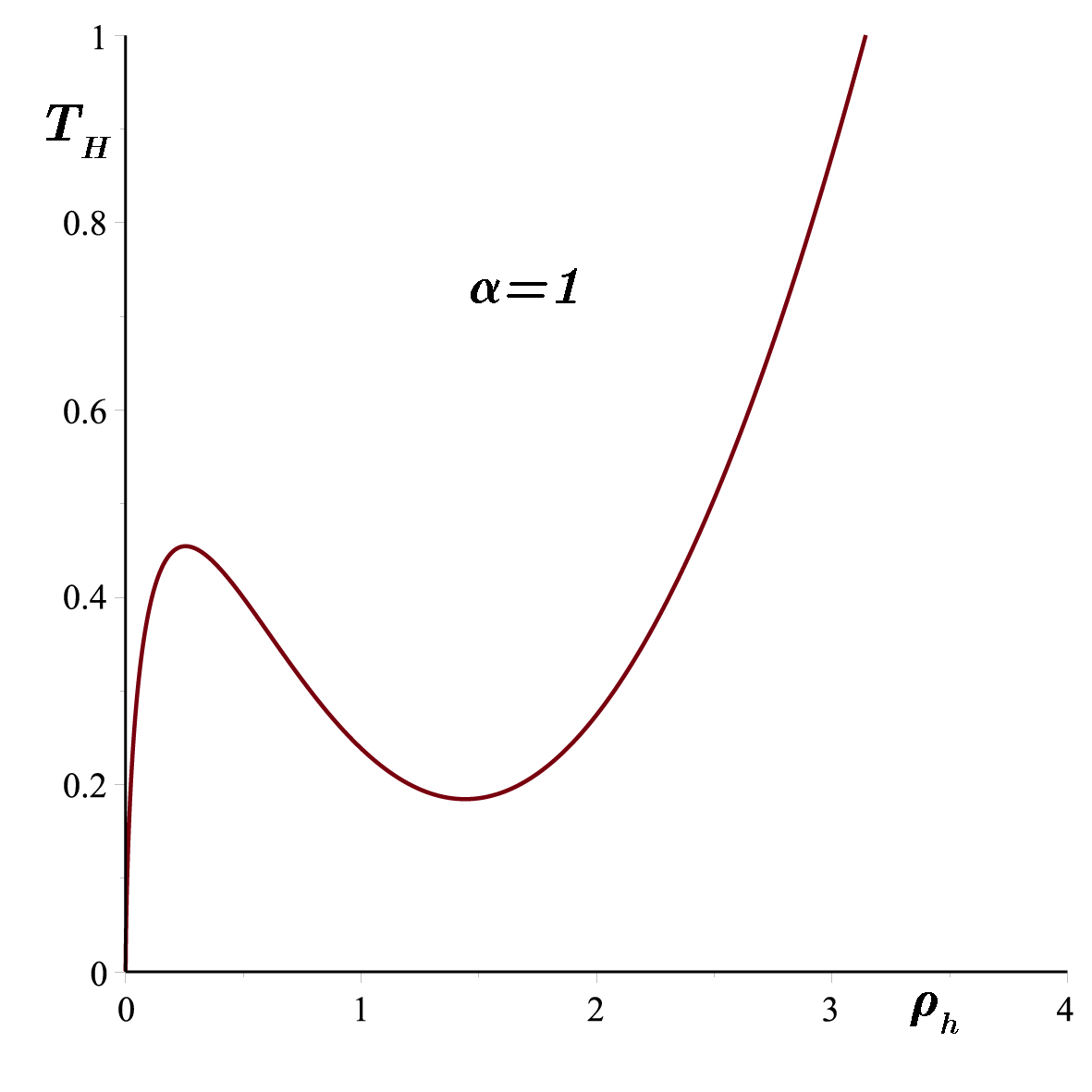}
\caption{Hawking temperature $T_{H}$ versus the event horizon radius $%
\protect\rho _{+}$ for $\protect\alpha =1.$ The minimum of the Hawking
temperature is observed (Eq. 47).}
\label{fig: 3}
\end{figure}
\begin{figure}[h]
\includegraphics[width=65mm,scale=0.7]{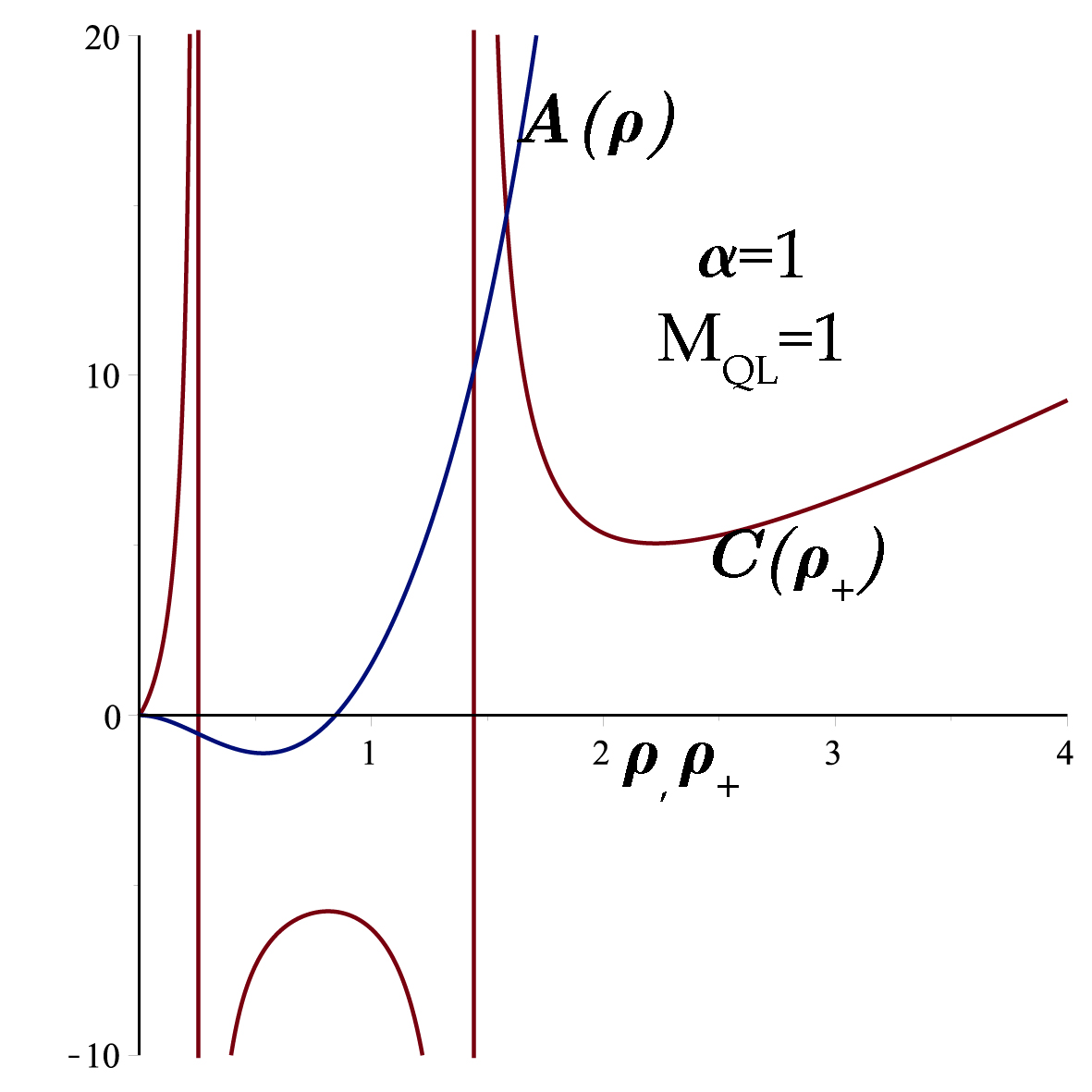}
\caption{Specific heat $C$ versus the event horizon radius $\protect\rho %
_{+} $ for $\protect\alpha =1.$ Also the metric function $A\left( \protect%
\rho \right) $ in terms of $\protect\rho $ for $\protect\alpha =1$ and $%
M_{QL}=1$ (Eqs. 43 and48).}
\label{fig: 4}
\end{figure}

\subsection{Thermodynamics}

The general solution found in the previous section is a two parameter
solution which are $M_{QL}$ and $\alpha .$ The other integration constants
are eliminated either by restriction or the fact that no new contribution
they provide, see for instance $C_{1}$ and $C_{2}$ in (19). We should also
admit that considering a relation between $H\left( r\right) $ and the scalar
field in the form given in (18) imposes restriction to our general solution.
As we stressed before, the solution may admit black hole with specific
values for the free parameters $M_{QL}$ and $\alpha $. In such a case, let's
assume that the metric function $A\left( \rho \right) $ in (31) admits an
event horizon located at $\rho =\rho _{+}.$ Using the standard definition of
the Hawking temperature%
\begin{equation}
T_{H}=\frac{1}{4\pi }\left. \sqrt{-g^{tt}g^{ij}g_{tt,i}g_{tt,j}}\right\vert
_{\rho =\rho _{+}}
\end{equation}%
one finds%
\begin{multline}
T_{H}=\frac{\left( \alpha ^{2}+2\right) \rho _{+}^{\left( \frac{4}{\alpha
^{2}}-2\right) }}{2\pi \alpha ^{2}}\times \\
\left\vert \frac{\left( \alpha ^{2}-2\right) ^{2}\ln ^{2}\rho _{+}+\alpha
^{2}\left( \alpha ^{2}-2\right) \ln \rho _{+}+\frac{\alpha ^{4}}{2}}{\left(
\alpha ^{2}-2\right) ^{3}}\right\vert .
\end{multline}%
Considering the entropy of the black hole at its horizon to be given by%
\begin{equation}
S=\frac{A_{+}}{4}
\end{equation}%
with $A_{+}$ the surface area of the horizon, the specific heat 
\begin{equation}
C=T_{H}\left( \frac{\partial S}{\partial T_{H}}\right)
\end{equation}%
yields%
\begin{multline}
C=-\frac{\pi \alpha ^{2}\rho _{+}}{2\left( \alpha ^{2}-2\right) }\times \\
\frac{\left( \alpha ^{2}-2\right) ^{2}\ln ^{2}\rho _{+}+\alpha ^{2}\left(
\alpha ^{2}-2\right) \ln \rho _{+}+\frac{\alpha ^{4}}{2}}{\left( \alpha
^{2}-2\right) ^{2}\ln ^{2}\rho _{+}-\alpha ^{2}\left( \alpha ^{2}-2\right)
\ln \rho _{+}-\frac{\alpha ^{4}}{2}}.
\end{multline}%
We note that, thermodynamically the black hole is locally stable if $C$ is
positive. Depending on the value of $\alpha $ and the radius of the horizon $%
\rho _{+},$ we may find stable $C>0$ or unstable $C<0$ black holes. This
suggests that the value of $M_{QL}$ which contributes to the radius of the
horizon, may be very crucial. In the case $\alpha =1$ in the following
section we shall give an example.

\subsection{Specific solution for $\protect\alpha =1$}

In this part we give the explicit solution for $\alpha =1.$ The metric
function and the potential become 
\begin{equation}
A\left( \rho \right) =-12M_{QL}\rho ^{2}-9\rho \left( \ln ^{2}\rho -2\ln
\rho +\frac{3}{2}\right) ,
\end{equation}%
and%
\begin{equation}
V\left( \phi \right) =-\left( 1-\phi +\phi ^{2}\right) e^{-\phi }
\end{equation}%
with%
\begin{equation}
\phi =2\ln \rho .
\end{equation}%
Therefore the line element takes the form%
\begin{equation}
ds^{2}=-A\left( \rho \right) dt^{2}+\frac{9\rho ^{4}}{A(\rho )}d\rho
^{2}+\rho ^{2}d\theta ^{2}.
\end{equation}%
The Hawking temperature is given by%
\begin{equation}
T_{H}=\left\vert \frac{3\left( 2\ln ^{2}\rho _{+}-2\ln \rho _{+}+1\right)
\rho _{+}}{4\pi }\right\vert ,
\end{equation}%
and the specific heat reads 
\begin{equation}
C=\frac{\pi \rho _{+}\left( \ln ^{2}\rho _{+}-\ln \rho _{+}+\frac{1}{2}%
\right) }{2\left( \ln ^{2}\rho _{+}+\ln \rho _{+}-\frac{1}{2}\right) }.
\end{equation}%
Fig. 1 is a plot of the metric function in terms of $\rho $ for $M_{QL}=%
\frac{1}{12},$ and $\alpha =1.$ For the same parameter values, in Fig. 2 we
plot the potential $V$ in terms of $\phi $ which clearly shows a local
minimum considered as the stability point of the field. In Fig. 3 we plot
the Hawking temperature $T_{H}$ given by Eq. (47). In this figure we observe
that a minimum temperature at certain horizon occurs. This horizon radius
can be considered as the minimum energy state of the black hole which is
more likely to admit a stable black hole. In Fig. 4, the specific heat $C$
and the metric function are displayed in terms of the event horizon $\rho
_{+}$ and $\rho $ respectively for $\alpha =1.$ Let's comment on this figure
that as the horizon of this specific black hole is located in the region
where the specific heat is negative, this black hole is not stable. The only
parameter that can be changed to shift the horizon into the region with
positive specific heat is the mass of the black hole i.e., $M_{QL}.$
Therefore increasing $M_{QL}$ causes the horizon to be larger and
consequently\ one obtains a positive $C$ and a stable black hole.

\subsection{A comparison with a singlet scalar field}

For an analytical comparison between the doublet scalar field and the
singlet scalar field one needs to consider in (1)%
\begin{equation}
\mathcal{L}_{field}=-\frac{1}{2}\left( \nabla \phi \right) ^{2}-V\left( \phi
\right) 
\end{equation}%
in which $\phi =\phi \left( r\right) $ is just a scalar field and $V\left(
\phi \right) $ is given by (5). Having the line element (6), we find%
\begin{equation}
\mathcal{L}_{field}=-\frac{A}{2}\phi ^{\prime 2}-V\left( \phi \right) 
\end{equation}%
and the scalar field equation is found to be%
\begin{equation}
\phi ^{\prime \prime }+\frac{\left( AH\right) ^{\prime }}{AH}\phi ^{\prime }-%
\frac{V^{\prime }\left( \phi \right) }{A}=0.
\end{equation}%
Einstein's equations (9) with 
\begin{equation}
T_{\mu }^{\nu }=\frac{1}{2}\left( \partial _{\mu }\phi \partial ^{\nu }\phi -%
\frac{1}{2}\left( \nabla \phi \right) ^{2}\delta _{\mu }^{\nu }\right) -%
\frac{1}{2}V\left( \phi \right) \delta _{\mu }^{\nu }
\end{equation}%
explicitly read%
\begin{equation}
\frac{2H^{\prime \prime }A+A^{\prime }H^{\prime }}{H}+\frac{1}{2}A\phi
^{\prime 2}+V=0,
\end{equation}%
\begin{equation}
\frac{A^{\prime }H^{\prime }}{H}-\frac{1}{2}A\phi ^{\prime 2}+V=0
\end{equation}%
and%
\begin{equation}
A^{\prime \prime }+\frac{1}{2}A\phi ^{\prime 2}+V=0.
\end{equation}%
Combining the Einstein's first two equations admits the same equation as
(17) and the ansatz 
\begin{equation}
H=e^{\mu \phi }
\end{equation}%
in which $\mu $ is a new constant parameter to be found, reveals a solution
for $\phi $ given by%
\begin{equation}
\phi \left( r\right) =\frac{2\mu }{2\mu ^{2}+1}\ln r.
\end{equation}%
Having the first Einstein's equation solved one finds $A\left( r\right) $
and making the other equations satisfied imposes $\xi _{1}=\xi _{2}=0$.
Finally the closed form of the general solution to the field equations is
given by%
\begin{equation}
A\left( r\right) =C_{1}r^{\frac{4\mu ^{2}}{2\mu ^{2}+1}}+C_{2}r^{\frac{1}{%
2\mu ^{2}+1}}
\end{equation}%
in which $C_{1}$ and $C_{2}$ are two integration constants. The final form
of the potential, however, becomes%
\begin{equation}
V=-\frac{2C_{1}\mu ^{2}\left( 4\mu ^{2}-1\right) }{\left( 2\mu ^{2}+1\right) 
}e^{-\phi /\mu },
\end{equation}%
which indicates that $\mu =\frac{1}{\alpha }$ and $V_{0}=-\frac{2C_{1}\mu
^{2}\left( 4\mu ^{2}-1\right) }{\left( 2\mu ^{2}+1\right) }$ when it is
compared with (5). In terms of the parameters introduced in the potential
function one may write the line element as%
\begin{multline}
ds^{2}=-\left( \frac{V_{0}\left( 2+\alpha ^{2}\right) ^{2}}{2\left( \alpha
^{2}-4\right) }r^{\frac{4}{2+\alpha ^{2}}}+C_{2}r^{\frac{\alpha ^{2}}{%
2+\alpha ^{2}}}\right) dt^{2}+ \\
\frac{dr^{2}}{\frac{V_{0}\left( 2+\alpha ^{2}\right) ^{2}}{2\left( \alpha
^{2}-4\right) }r^{\frac{4}{2+\alpha ^{2}}}+C_{2}r^{\frac{\alpha ^{2}}{%
2+\alpha ^{2}}}}+r^{\frac{4}{2+\alpha ^{2}}}d\theta ^{2}.
\end{multline}%
Let's note that $\alpha =2$ must be excluded and the corresponding specific
solution is given by%
\begin{equation}
A\left( r\right) =r^{2/3}\left( C-3V_{0}\ln r\right) 
\end{equation}%
with $H,$ $V$ and $\phi $ the same as the general $\alpha $ with $\alpha =2$
($\mu =\frac{1}{2}$).

For the specific value of $\mu =\frac{1}{\sqrt{2}}$ ($\alpha =\sqrt{2}$) we
find the line element to be 
\begin{equation}
ds^{2}=-\left( C_{2}\sqrt{r}-4V_{0}r\right) dt^{2}+\frac{dr^{2}}{C_{2}\sqrt{r%
}-4V_{0}r}+rd\theta ^{2}
\end{equation}%
which after the transformation $r=\rho ^{2}$ becomes%
\begin{equation}
ds^{2}=-\rho \left( C_{2}-4V_{0}\rho \right) dt^{2}+\frac{4\rho d\rho ^{2}}{%
C_{2}-4V_{0}\rho }+\rho ^{2}d\theta ^{2}.
\end{equation}%
The scalar field and the potential read as%
\begin{equation}
\phi =\sqrt{2}\ln \rho
\end{equation}%
and%
\begin{equation}
V=\frac{V_{0}}{\rho ^{2}}.
\end{equation}%
The solutions given by (56)-(59) represent three parameters solutions which
are $\alpha ,$ $C_{2}$ and $V_{0}.$ With proper choice of parameters the
general solution (i.e., (58) or (61)) admits black hole solution. For
instance, in the case $\alpha =\sqrt{2},$ if we assume that both $C_{2}$ and 
$V_{0}$ are negative the solution is a black hole with quasilocal mass given
by the BY formalism (34) as%
\begin{equation}
M_{QL}=\frac{\left\vert C_{2}\right\vert }{8}.
\end{equation}%
The differences between the singlet and doublet field equations as well as
solutions are very clear. Finally let us add that by setting $C_{2}=0$ and $%
V_{0}=-1,$ the line element (63) becomes%
\begin{equation}
ds^{2}=-\rho ^{2}d\bar{t}^{2}+d\rho ^{2}+\rho ^{2}d\theta ^{2}
\end{equation}%
where $\bar{t}=2t.$ This solution was found in \cite{41}.

\section{An extension to $3+1-$dimensions}

In $3+1-$dimensions the action reads as 
\begin{equation}
S=\int d^{4}x\sqrt{-g}\left( R+\mathcal{L}_{field}\right)
\end{equation}%
in which $\mathcal{L}_{field}$ is given by (2), however, the components of
the triplet scalar potential as source are given by 
\begin{eqnarray}
\phi ^{1} &=&\phi \left( r\right) \sin \theta \cos \varphi \\
\phi ^{2} &=&\phi \left( r\right) \sin \theta \sin \varphi \\
\phi ^{3} &=&\phi \left( r\right) \cos \theta
\end{eqnarray}%
with its modulus given as in (4) with $a=1,2,3$. The self interacting
potential is considered as $V\left( \phi \right) $ while the spherically
symmetric line element is chosen to be 
\begin{equation}
ds^{2}=-A\left( r\right) dt^{2}+\frac{1}{A(r)}dr^{2}+H\left( r\right)
^{2}\left( d\theta ^{2}+\sin ^{2}\theta d\varphi ^{2}\right) .
\end{equation}%
Similar to the three dimensional case, the field equations are given by the
variation of the action with respect to $\phi \left( r\right) $ which yields%
\begin{equation}
\left( A\phi ^{\prime }H^{2}\right) ^{\prime }-2\phi =H^{2}V^{\prime }\left(
\phi \right)
\end{equation}%
and with respect to the metric tensor which gives the Einstein's equations
with the energy-momentum tensor as in (10). These Einstein's equations may
be combined and in their simplest form they become%
\begin{equation}
\left( A^{\prime }H^{2}\right) ^{\prime }+H^{2}V=0,
\end{equation}%
\begin{equation}
4H^{\prime \prime }+H\phi ^{\prime 2}=0,
\end{equation}%
\begin{equation}
A\left( H^{2}\right) ^{\prime \prime }-H^{2}A^{\prime \prime }+\left( \phi
^{2}-2\right) =0,
\end{equation}%
and%
\begin{equation}
2\left( A^{\prime }HH^{\prime }+AH^{\prime 2}-1\right) -\left( \frac{%
AH^{2}\phi ^{\prime 2}}{2}-\phi ^{2}-H^{2}V\right) =0.
\end{equation}

\section{Solution to the field equations in $3+1-$dimensions}

One can check that the following set of functions for $\phi \left( r\right)
, $ $H\left( r\right) ,$ $A\left( r\right) $ and $V\left( \phi \right) $
satisfy all field equations,%
\begin{equation}
\phi =\frac{2\alpha }{\alpha ^{2}+1}\ln r,
\end{equation}%
\begin{equation}
H\left( r\right) =r^{\frac{\alpha ^{2}}{\alpha ^{2}+1}},
\end{equation}%
\begin{multline}
A\left( r\right) =-\frac{\Lambda }{3}r^{\frac{2\alpha ^{2}}{1+\alpha ^{2}}}-%
\frac{2M}{r^{\frac{\alpha ^{2}-1}{\alpha ^{2}+1}}}-\frac{2\alpha ^{2}}{%
\left( \alpha ^{4}-1\right) }r^{\frac{2}{\alpha ^{2}+1}}\times \\
\left( \ln ^{2}r-\frac{\alpha ^{2}-3}{\alpha ^{2}-1}\ln r-\frac{\alpha
^{8}-3\alpha ^{6}+4\alpha ^{4}-7\alpha ^{2}+1}{2\alpha ^{2}\left( \alpha
^{2}-1\right) ^{2}}\right)
\end{multline}%
and 
\begin{multline}
V\left( \phi \right) =\frac{2}{3}\frac{\alpha ^{2}\left( 3\alpha
^{2}-1\right) \Lambda }{\left( \alpha ^{2}+1\right) ^{2}}e^{-\frac{\phi }{%
\alpha }}+\frac{2\left( 2\alpha ^{2}-1\right) }{\left( \alpha ^{2}-1\right)
^{3}}\times \\
\left( 1+\frac{\alpha ^{3}\left( \alpha ^{2}-1\right) }{\left( 2\alpha
^{2}-1\right) }\phi +\frac{\left( \alpha ^{2}-1\right) ^{2}}{2\left( 2\alpha
^{2}-1\right) }\phi ^{2}\right) e^{-\alpha \phi }.
\end{multline}%
Here, $\alpha $ is a free real parameter such that $\alpha \in \left[
0,\infty \right) -\left\{ 1,\frac{1}{\sqrt{2}}\right\} $ while $M$ and $%
\Lambda $ are two integration constants corresponding to the mass of the
black hole and the cosmological constant. Furthermore, in the limit $\alpha
\rightarrow \infty $ one finds%
\begin{eqnarray}
\phi &=&0, \\
H\left( r\right) &=&r, \\
A\left( r\right) &=&1-\frac{\Lambda }{3}r^{2}-\frac{2M}{r}
\end{eqnarray}%
and%
\begin{equation}
V\left( \phi \right) =2\Lambda
\end{equation}%
which is the (anti) de-Sitter Schwarzschild black hole solution. Let's add
also that the limit $\alpha =0$ gives the Bertotti-Kasner spacetime \cite{44}%
. Two particular cases corresponding to $\alpha =1$ and $\alpha =\frac{1}{%
\sqrt{2}}$ that were excluded above will be considered separately in the
sequel. Before that we would like to look at the general solution more
closely. First we apply the following transformation%
\begin{equation}
r^{\frac{\alpha ^{2}}{\alpha ^{2}+1}}=x\text{ }
\end{equation}%
which yields%
\begin{equation}
ds^{2}=-A\left( x\right) dt^{2}+\frac{\left( 1+\frac{1}{\alpha ^{2}}\right)
^{2}x^{\frac{2}{\alpha ^{2}}}dx^{2}}{A\left( x\right) }+x^{2}d\Omega ^{2}
\end{equation}%
in which%
\begin{multline}
A\left( x\right) =-\frac{\Lambda }{3}x^{2}-\frac{2M}{x^{1-\frac{1}{\alpha
^{2}}}}-\frac{2\left( \alpha ^{2}+1\right) ^{2}x^{\frac{2}{\alpha ^{2}}}}{%
\alpha ^{2}\left( \alpha ^{4}-1\right) }\times \\
\left( \ln ^{2}x-\frac{\alpha ^{2}\left( \alpha ^{2}-3\right) \ln x}{\alpha
^{4}-1}\right. \\
-\left. \frac{\left( \alpha ^{8}-3\alpha ^{6}+4\alpha ^{4}-7\alpha
^{2}+1\right) \alpha ^{2}}{2\left( \alpha ^{4}-1\right) ^{2}}\right) .
\end{multline}%
What we observe is that $\Lambda $ is still an effective cosmological
constant and the spacetime admits black holes. Once more we apply the BY
formalism to find the quasilocal mass of the possible central black hole. To
do so we set $F\left( x_{B}\right) ^{2}=A\left( x_{B}\right) $ and $G\left(
x_{B}\right) ^{2}=\frac{A\left( x_{B}\right) }{\left( 1+\frac{1}{\alpha ^{2}}%
\right) ^{2}x_{B}^{\frac{2}{\alpha ^{2}}}}$ in Eq. (34) which results in%
\begin{equation}
M_{QL}=\frac{M}{\left( 1+\frac{1}{\alpha ^{2}}\right) }.
\end{equation}%
Clearly at the limit of $\alpha \rightarrow \infty $ the quasilocal mass
reduces into ADM mass of the central Schwarzschild black hole. To complete
our discussion on the solution given above, we add that the solution is
singular only at the origin and it diverges as fast as $\frac{\ln ^{3}x}{%
x^{4}}.$

\subsection{Thermodynamics}

Similar to the black hole solution in $2+1-$dimensions, here we determine
some basic thermodynamic properties of the black hole solution given in (87)
and (88). The Hawking temperature is found to be%
\begin{multline}
T_{H}=\frac{x_{+}^{-\left( 1+\frac{1}{\alpha ^{2}}\right) }}{4\pi \alpha
^{2}\left( 1+\alpha ^{2}\right) }\left\vert \alpha ^{2}\Lambda \left( \alpha
^{2}-\frac{1}{3}\right) x_{+}^{2}+\left( \alpha ^{2}+1\right) ^{2}x_{+}^{%
\frac{2}{\alpha ^{2}}}\times \right. \\
\left. \frac{2\left( \alpha ^{2}-1\right) ^{2}\ln ^{2}x_{+}+2\alpha
^{2}\left( \alpha ^{2}-1\right) \ln x_{+}-\alpha ^{2}\left( \alpha
^{4}-3\alpha ^{2}+1\right) }{\left( \alpha ^{2}-1\right) ^{3}}\right\vert
\end{multline}%
in which $x_{+}$ is the radius of the event horizon. Having the entropy of
the black hole to be%
\begin{equation*}
S=\pi x_{+}^{2}
\end{equation*}%
and the definition of the specific heat (41), we determine%
\begin{multline}
C=\frac{2\pi \alpha ^{2}x_{+}^{2}}{\left( \alpha ^{2}-1\right) }\times \\
\frac{\alpha ^{2}\Lambda \left( \alpha ^{2}-\frac{1}{3}\right) \left( \alpha
^{2}-1\right) ^{3}x_{+}^{2}-\Gamma _{1}\left( \alpha ^{2}+1\right)
^{2}x_{+}^{\frac{2}{\alpha ^{2}}}}{\alpha ^{2}\Lambda \left( \alpha ^{2}-%
\frac{1}{3}\right) \left( \alpha ^{2}-1\right) ^{3}x_{+}^{2}+\Gamma
_{2}\left( \alpha ^{2}+1\right) ^{2}x_{+}^{\frac{2}{\alpha ^{2}}}},
\end{multline}%
where%
\begin{multline}
\Gamma _{1}=-2\left( \alpha ^{2}-1\right) ^{2}\ln ^{2}x_{+}- \\
2\alpha ^{2}\left( \alpha ^{2}-1\right) \ln x_{+}+\alpha ^{2}\left( \alpha
^{4}-3\alpha ^{2}+1\right)
\end{multline}%
and%
\begin{multline}
\Gamma _{2}=-2\left( \alpha ^{2}-1\right) ^{2}\ln ^{2}x_{+}+ \\
2\alpha ^{2}\left( \alpha ^{2}-1\right) \ln x_{+}+\alpha ^{2}\left( \alpha
^{4}-3\alpha ^{2}+1\right) .
\end{multline}%
We should add that, as of the $2+1-$dimensions, for a specific radius for
the event horizon, depending on the sign of the heat capacity, the black
hole, thermodynamically, is locally stable or unstable. In Fig. 5 we plot
the specific heat $C\left( x_{+}\right) $ and the metric function $A\left(
x\right) $ in terms of $x_{+}$ and $x$ respectively for $\alpha =2,$ $%
\Lambda =-4$ and $M_{QL}=1.$ Since $C$ is not a function of $M_{QL},$
changing the value of $M_{QL}$ does not alter $C$ however it changes the
radius of the horizon. As in Fig. 5, for $M_{QL}=1$ the horizon located in
the region where $C>0$ and as a result the black hole is stable. Decreasing $%
M_{QL}$ causes the horizon fall in the region where $C<0$ and the black hole
is no longer stable.

In Fig. 6 we plot the Hawking temperature $T_{H}\left( x_{+}\right) $ and
the metric function $A\left( x\right) $ in terms of $x_{+}$ and $x$
respectively for $\alpha =2,$ $\Lambda =-4$ and $M_{QL}=1.$ The Hawking
temperature admits a zero and a local minimum.

\begin{figure}[h]
\includegraphics[width=65mm,scale=0.7]{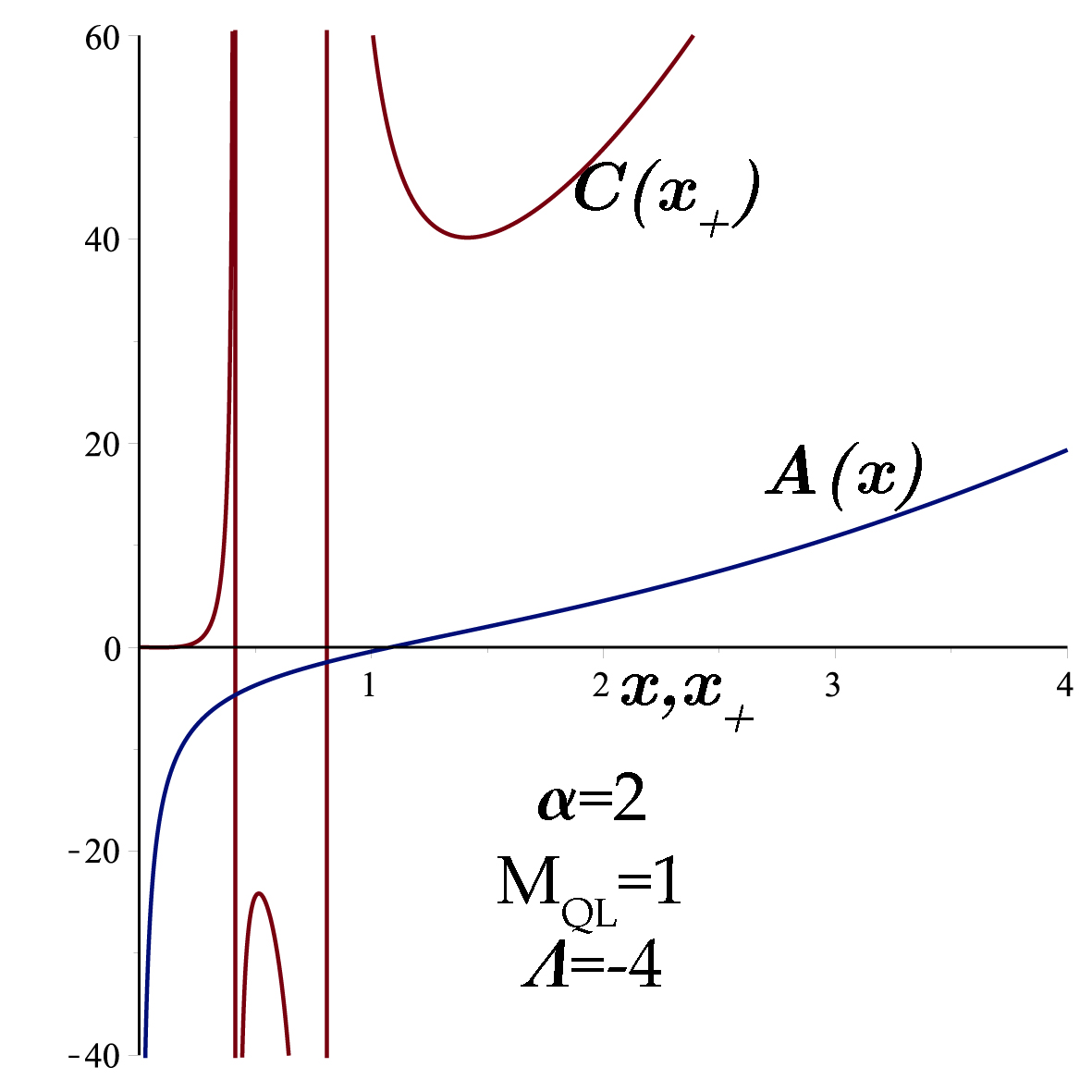}
\caption{$C\left( x_{+}\right) $ and the metric function $A\left( x\right) $
in terms of $x_{+}$ and $x$ respectively for $\protect\alpha =2,$ $\Lambda
=-4$ and $M_{QL}=1$ (Eqs. 88 and 91).}
\label{fig: 5}
\end{figure}
\begin{figure}[h]
\includegraphics[width=65mm,scale=0.7]{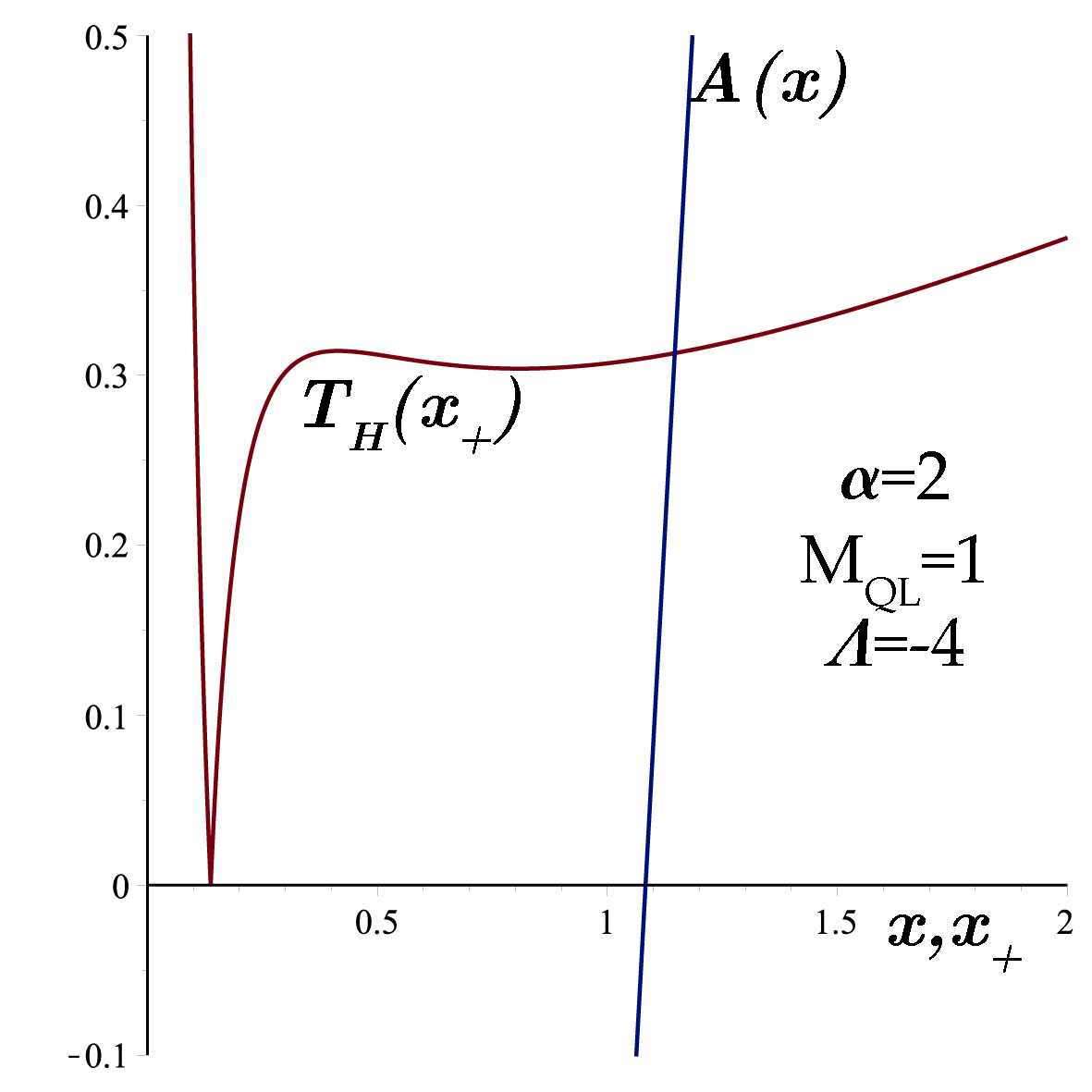}
\caption{$T_{H}\left( x_{+}\right) $ and the metric function $A\left(
x\right) $ in terms of $x_{+}$ and $x$ respectively for $\protect\alpha =2,$ 
$\Lambda =-4$ and $M_{QL}=1$ (Eqs. 88 and 90).}
\label{fig: 6}
\end{figure}

\subsection{$\protect\alpha =1$}

The solution for $\alpha =1$ which is obtained separately apart from the
solution (78-81) becomes%
\begin{equation}
\phi =\ln r,
\end{equation}%
\begin{equation}
H\left( r\right) =\sqrt{r},
\end{equation}%
\begin{equation}
A\left( r\right) =C_{2}+C_{1}r-r\ln ^{2}r+\frac{r}{3}\ln ^{3}r
\end{equation}%
and 
\begin{equation}
V\left( \phi \right) =-\frac{1}{3}\left( 3C_{1}-6-6\phi +3\phi ^{2}+\phi
^{3}\right) e^{-\phi }
\end{equation}%
in which $C_{1}$ and $C_{2}$ are two integration constants and the metric
implies a non-asymptotically flat black hole solution. A coordinate
transformation of the form $x=\sqrt{r}$ transforms the line element (69) into%
\begin{equation}
ds^{2}=-A\left( x\right) dt^{2}+\frac{4x^{2}}{A\left( x\right) }%
dx^{2}+x^{2}d\Omega ^{2}
\end{equation}%
with%
\begin{equation}
A\left( x\right) =-4M_{QL}+C_{1}x^{2}-4x^{2}\ln ^{2}x+\frac{8x^{2}}{3}\ln
^{3}x.
\end{equation}%
Note that to find the quasilocal mass as $M_{QL}=\frac{-C_{2}}{4}$ we
applied the BY formalism. Let's add that the constant $C_{1}$ is an
effective cosmological constant. In Fig. 7 we plot $V\left( \phi \right) $
versus $\phi $ for $C_{1}=0.$ In this figure we observe a kind of modified
Mexican hat potential with the left minimum much deeper. In Fig. 8 we depict 
$A\left( x\right) $ versus $x$ for $C_{1}=0$ and $M_{QL}=1$ respectively. 
\begin{figure}[h]
\includegraphics[width=65mm,scale=0.7]{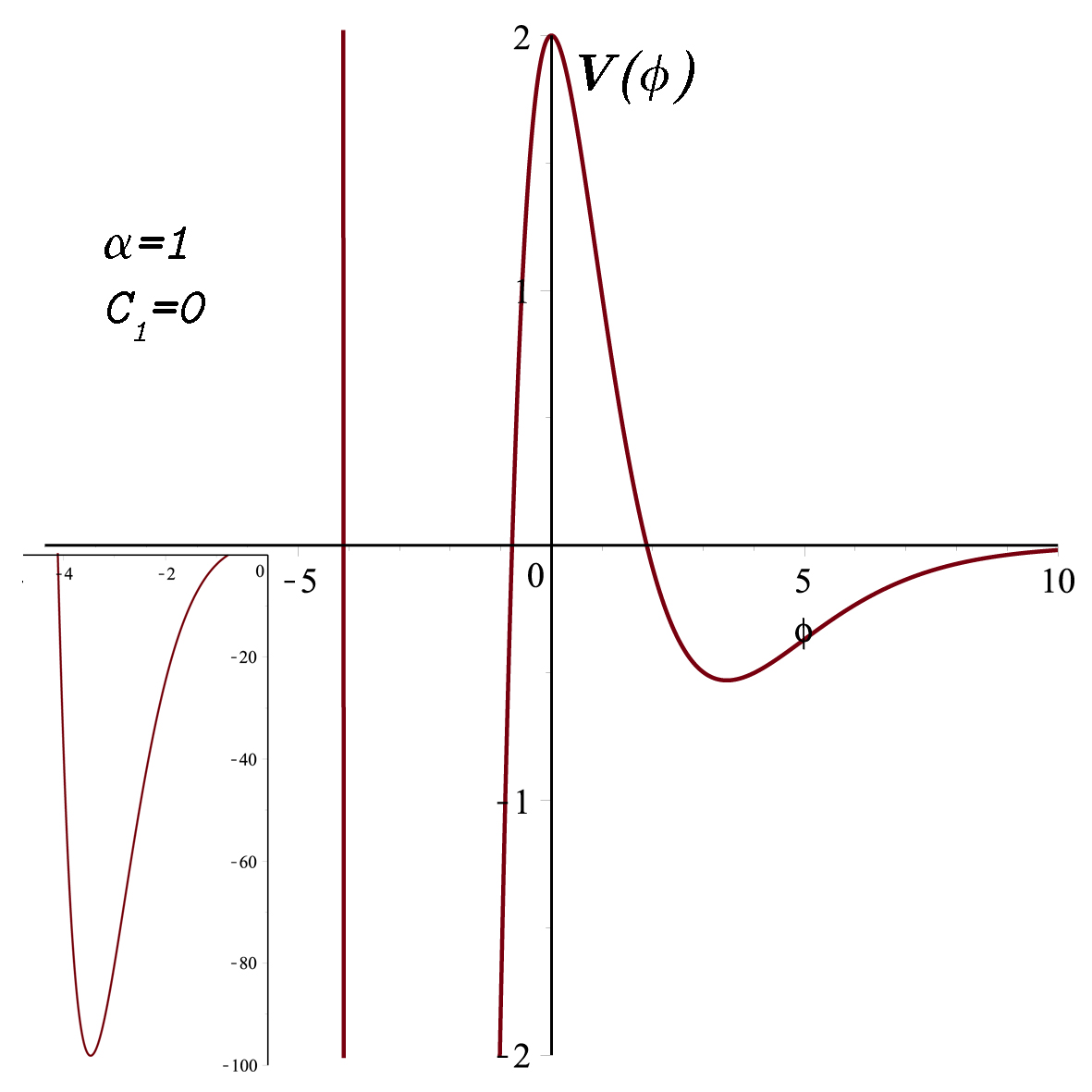}
\caption{$V\left( \protect\phi \right) $ versus $\protect\phi $ for $C_{1}=0$
and $\protect\alpha =1$ in $3+1$-dimensions i.e., Eq. (97).}
\label{fig: 7}
\end{figure}

\begin{figure}[h]
\includegraphics[width=65mm,scale=0.7]{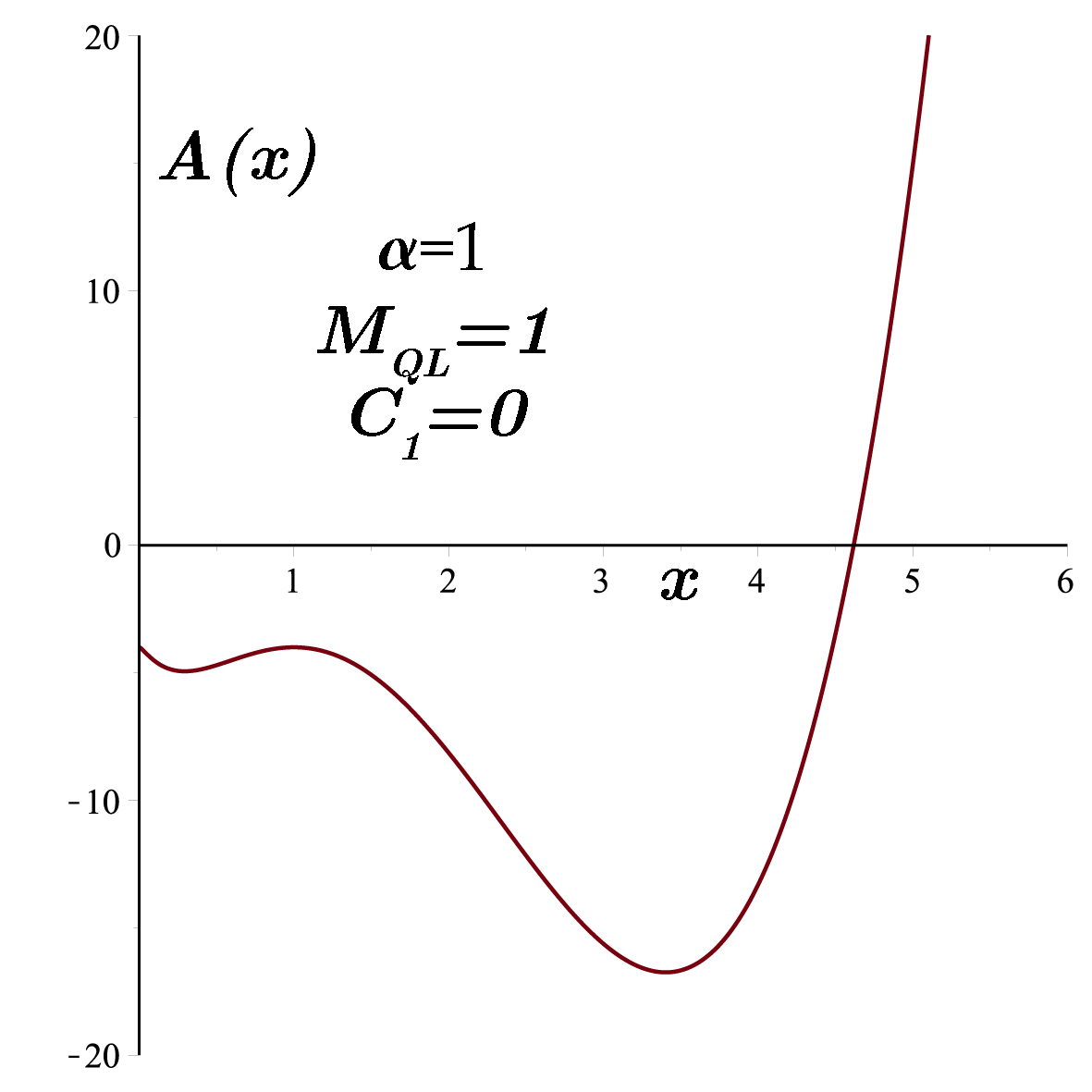}
\caption{$A\left( x\right) $ versus $x$ for $C_{1}=0,$ $M_{QL}=1$ and $%
\protect\alpha =1$ in $3+1$-dimensions i.e., Eq. (99).}
\label{fig: 8}
\end{figure}
\begin{figure}[h]
\includegraphics[width=65mm,scale=0.7]{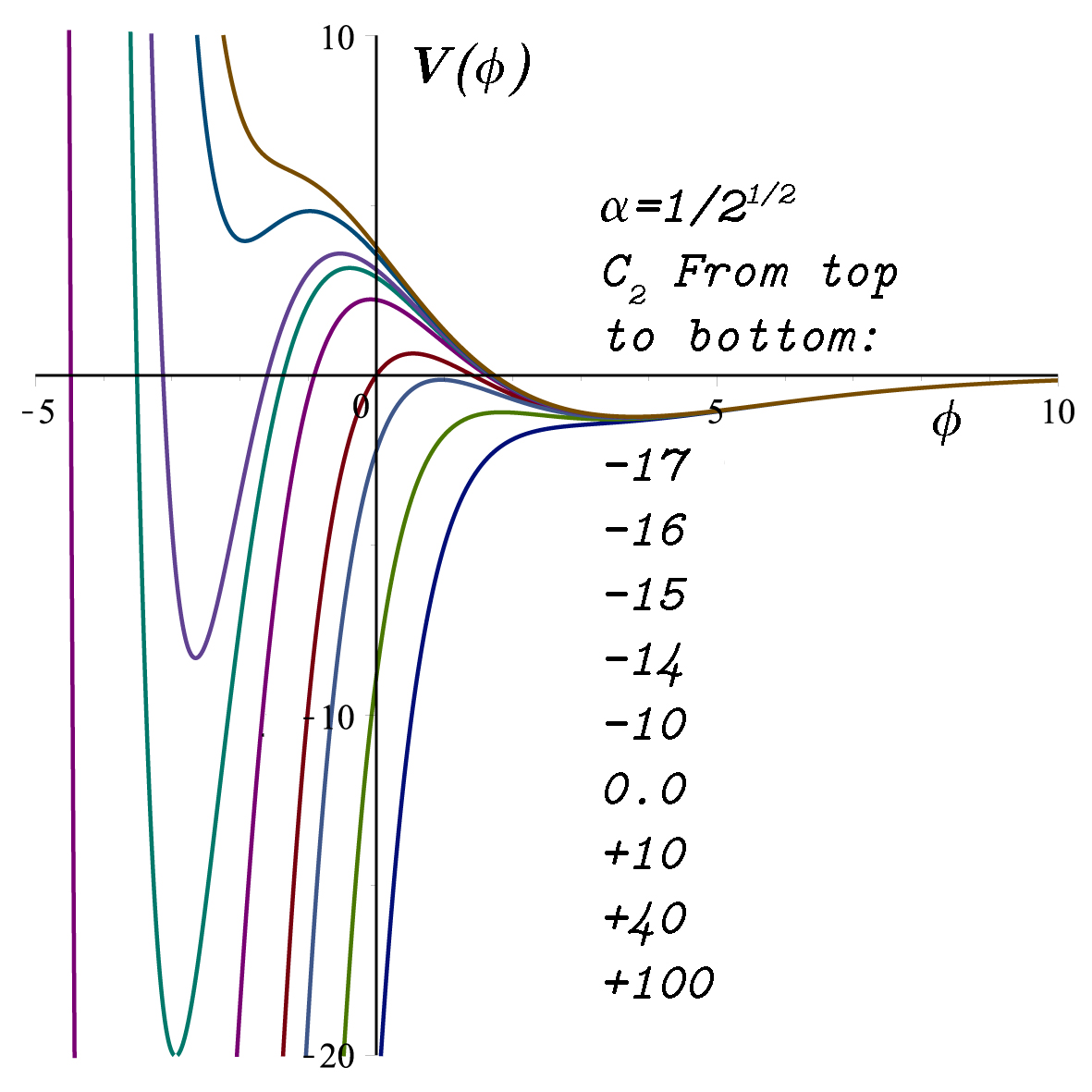}
\caption{$V\left( \protect\phi \right) $ versus $\protect\phi $ for various
values for $C_{2}$ and $\protect\alpha =\frac{1}{\protect\sqrt{2}}$ in $3+1$%
-dimensions i.e., Eq. (103).}
\label{fig: 9}
\end{figure}
\begin{figure}[h]
\includegraphics[width=65mm,scale=0.7]{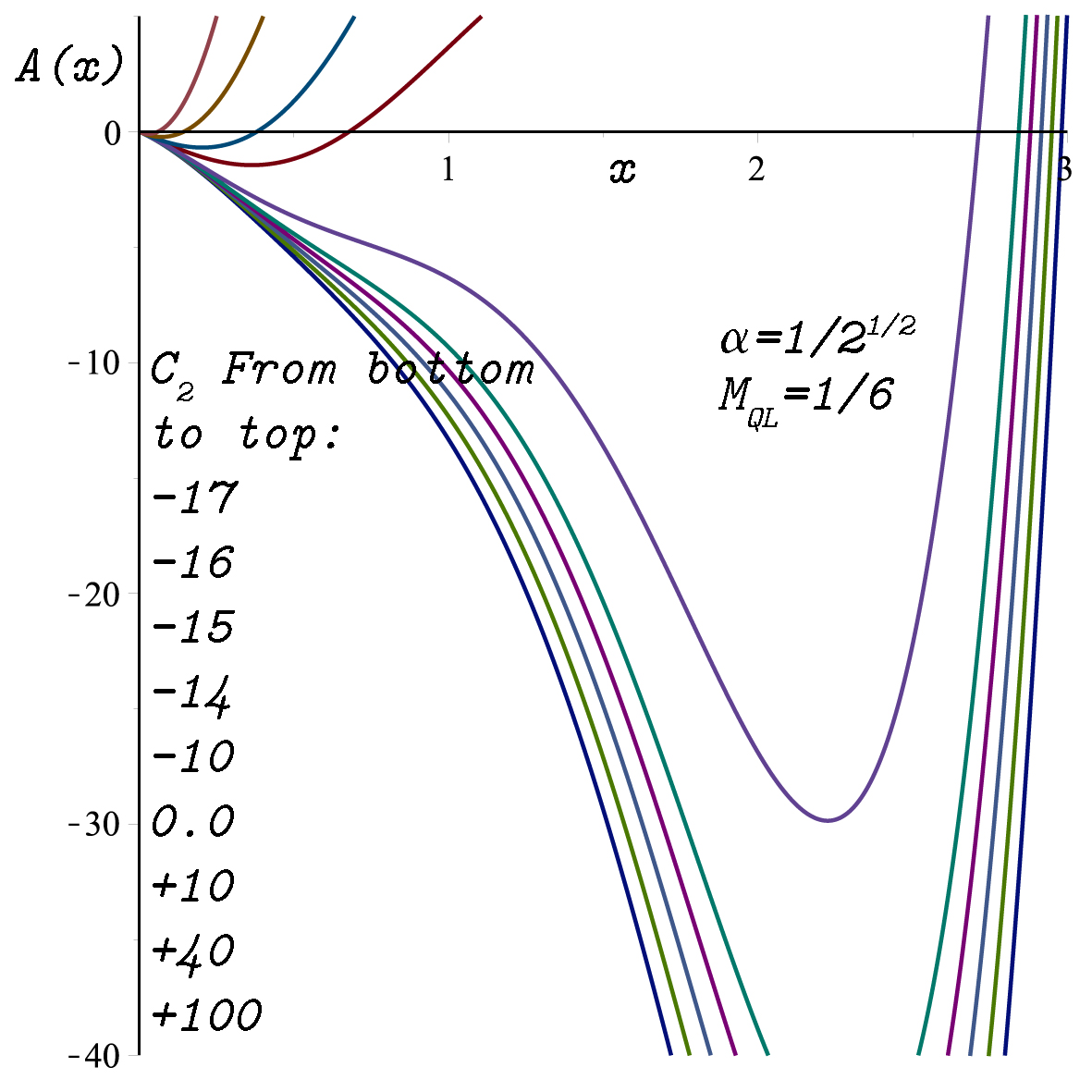}
\caption{$A\left( x\right) $ versus $x$ for various values for $C_{2}$, $%
M_{QL}=\frac{1}{6}$ and $\protect\alpha =\frac{1}{\protect\sqrt{2}}$ in $3+1$%
-dimensions i.e., Eq. (106).}
\label{fig: 10}
\end{figure}

\subsection{$\protect\alpha =\frac{1}{\protect\sqrt{2}}$}

The solution for $\alpha =\frac{1}{\sqrt{2}}$ yields%
\begin{equation}
\phi =\frac{2\sqrt{2}}{3}\ln r,
\end{equation}%
\begin{equation}
H\left( r\right) =\sqrt[3]{r},
\end{equation}%
\begin{equation}
A\left( r\right) =r^{\frac{2}{3}}C_{2}+C_{1}r^{\frac{1}{3}}+\frac{1}{3}r^{%
\frac{4}{3}}\left( 29-20\ln r+4\ln ^{2}r\right)
\end{equation}%
and 
\begin{equation}
V\left( \phi \right) =-\frac{2}{9}C_{2}e^{-\sqrt{2}\phi }+2\phi \left( \sqrt{%
2}-\phi \right) e^{-\frac{\sqrt{2}}{2}\phi }
\end{equation}%
in which $C_{1}$ and $C_{2}$ are two integration constants. Once more we
transform our solution by applying the following coordinate transformation 
\begin{equation}
r=x^{3}
\end{equation}%
which modifies the line element into the form%
\begin{equation}
ds^{2}=-A\left( x\right) dt^{2}+\frac{9x^{4}}{A\left( x\right) }%
dx^{2}+x^{2}d\Omega ^{2}
\end{equation}%
where%
\begin{equation}
A\left( x\right) =x^{2}C_{2}-6M_{QL}x+\frac{1}{3}x^{4}\left( 29-60\ln
x+36\ln ^{2}x\right) .
\end{equation}%
We add that from the BY formalism we found the quasilocal mass of the black
hole solution as $M_{QL}=-\frac{C_{1}}{6}.$ In Fig. 9 we depicted $V\left(
\phi \right) $ versus $\phi $ for various $C_{2}$ which is the effective
cosmological constant. We observe that for a negative $C_{2}$ there is at
most two local minima but for the positive values only one local minimum is
found. In Fig. 10 we plot the corresponding metric function for $M_{QL}=%
\frac{1}{6}$.

\section{Conclusion}

By employing doublet and triplet of scalars with a self-interacting
potential consisting only of their moduli we constructed classes of
non-asymptotically flat black holes in $2+1-$ and $3+1-$ dimensions. The
self-interacting potential consists of a polynomial term multiplied by a
Liouville term. The latter factor is the potential which plays the role to
dominate the asymptotic behaviors. Our model can be considered within the
context of field theoretic black holes. Some thermodynamical properties
including specific heat and quasilocal mass are given explicitly. The
simplest member of our model is naturally the case of a constant potential
term which corresponds to the cosmological constant. It is shown that the
potential admits local minimum apt for the construction of suitable field
theoretic black hole states. One distinguishing feature of our metric
function obtained from scalar multiplets is that it is a polynomial of a
mixture of radial function and its logarithms. It can be anticipated that
given the proper self-interacting potential our model of multiplets can be
extended to higher dimensions. \ Unless this has been worked out explicitly,
however, based only on $2+1$ / $3+1$ dimensions, it is hard to predict the
higher dimensional behaviors and formulate a general no-go theorem in the
presence of multiplet sources. In our restricted dimensions we obtained no
asymptotically flat regular black hole solutions for $V\left( \phi \right)
>0 $, which is in conform with the no-go theorem introduced in \cite{45,46}.

\bigskip


\begin{thebibliography}{99}
\bibitem{1} M. Ba\~{n}ados, C. Teitelboim, J. Zanelli, Phys. Rev. Lett. 
\textbf{69}, 1849 (1992).

\bibitem{2} M. Ba\~{n}ados, M. Henneaux, C. Teitelboim and J. Zanelli, Phys.
Rev. D \textbf{48}, 1506 (1993).

\bibitem{3} C. Martinez, C. Teitelboim and J. Zanelli, Phys. Rev. D \textbf{%
61,} 104013 (2000).

\bibitem{4} S. Carlip, \textit{Quantum Gravity in 2 + 1-Dimensions},
Cambridge University Press, 1998.

\bibitem{5} S. Carlip, Living Rev. Rel. \textbf{8}, 1 (2005).

\bibitem{6} S. H. Mazharimousavi and M. Halilsoy, Phys. Rev. D \textbf{92},
024040 (2015).

\bibitem{7} S. H. Mazharimousavi and M. Halilsoy, Eur. Phys. J. C \textbf{75}%
, 249 (2015).

\bibitem{8} A. L. Berkin and R. W. Hellings, Phys. Rev. D \textbf{49}, 6442
(1994).

\bibitem{9} V. Vardanyan and L. Amendola, Phys. Rev. D \textbf{92}, 024009
(2015).

\bibitem{10} M. Rainer and A. Zhuk, Phys. Rev. D \textbf{54}, 6186 (1996).

\bibitem{11} D. I. Kaiser, Phys. Rev. D \textbf{81}, 084044 (2010).

\bibitem{12} Y. Watanabe and J. White, Phys. Rev. D \textbf{92}, 023504
(2015).

\bibitem{13} J. White, M. Minamitsuji and M. Sasaki, JCAP \textbf{07, }039%
\textbf{\ }(2012).

\bibitem{14} K. Schutz, E. I. Sfakianakis and D. I. Kaiser, Phys. Rev. D 
\textbf{89}, 064044 (2014).

\bibitem{15} D. H. Lyth and A. Riotto, Phys. Rep. \textbf{314}, 1 (1999).

\bibitem{16} C. P. Burgess, Classical Quantum Gravity \textbf{24}, S795
(2007).

\bibitem{17} L. McAllister and E. Silverstein, Gen. Relativ. Gravit. \textbf{%
40}, 565 (2008).

\bibitem{18} D. Baumann and L. McAllister, Annu. Rev. Nucl. Part. Sci. 
\textbf{59}, 67 (2009).

\bibitem{19} A. Mazumdar and J. Rocher, Phys. Rep. \textbf{497}, 85 (2011).

\bibitem{20} M. Barriola and A. Vilenkin, Phys. Rev. Lett. \textbf{63}, 341
(1989).

\bibitem{21} A. Vilenkin, Phys. Rep. \textbf{121}, 263 (1985).

\bibitem{22} C. M. Chen, H. B. Cheng, X. Z. Li, X. H. Zhai, Class. Quantum
Gravity \textbf{13}, 701 (1996).

\bibitem{23} X. Z. Li, Commun. Theor. Phys. \textbf{28}, 101 (1997).

\bibitem{24} D. Harari and C. Lousto, Phys. Rev. D \textbf{42}, 2626 (1990).

\bibitem{25} O. Dando and R. Gregory, Class. Quantum Grav. \textbf{15}, 985
(1998).

\bibitem{26} A. Banerjee, A. Beesham, S. Chatterjee and A. A. Sen, Class.
Quantum Grav. \textbf{15}, 645 (1998).

\bibitem{27} T. H. Lee and B. J. Lee, Phys. Rev. D \textbf{69}, 127502
(2004).

\bibitem{28} R. M. Teixeira Filho and V. B. Bezerra, Phys. Rev. D \textbf{64}%
, 067502 (2001).

\bibitem{29} T. Tamaki and K-ichi Maeda, Phys. Rev. D \textbf{60}, 104049
(1999).

\bibitem{30} E. Hirschmann, Anzhong Wang, Y. Wu, Class. Quant. Grav. \textbf{%
21,} 1791 (2004).

\bibitem{31} E. Ay\'{o}n-Beato, A. Garcia, A. Macias, J. Perez-Sanchez,
Phys. Lett. B \textbf{495}, 164 (2000).

\bibitem{32} I. Z. Fisher, Z. Exp. Teor. Fiz. \textbf{18}, 636 (1948).

\bibitem{33} T. Kodama, Phys. Rev. D \textbf{18}, 3529 (1978).

\bibitem{34} T. Kodama, L. de Oliveira, F. Santos, Phys. Rev. D \textbf{19},
3576 (1979).

\bibitem{35} P. Baekler, E. Mielke, R. Hecht, F. Hehl, Nucl. Phys. B \textbf{%
288}, 800 (1987).

\bibitem{36} K. Schmoltzi, T. Sch\"{u}cker, Phys. Lett. A \textbf{161}, 212
(1991).

\bibitem{37} P. Jetzer, D. Scialom, Phys. Lett. A \textbf{169}, 12 (1992).

\bibitem{38} T. Torii, K. Maeda, and M. Narita, Phys. Rev. D \textbf{64},
044007 (2001).

\bibitem{39} E. Winstanley, Found. Phys. \textbf{33}, 111 (2003).

\bibitem{40} C. A. R. Herdeiro and E. Radu, Int. J. Mod. Phys. D \textbf{24,}
1542014 (2015).

\bibitem{41} H.-J. Schmidt and D. Singleton, Phys. Lett. B \textbf{721}, 294
(2013).

\bibitem{42} J. D. Brown and J. W. York, Phys. Rev. D \textbf{47}, 1407
(1993).

\bibitem{43} J. D. Brown, J. Creighton and R. B. Mann, Phys. Rev. D \textbf{%
50,} 6394 (1994).

\bibitem{44} W. Rindler, Phys. Lett. A, \textbf{245}, 363 (1998).

\bibitem{45} K. A. Bronnikov and G.N. Shikin, Grav.Cosmol. \textbf{8}, 107
(2002).

\bibitem{46} K. A. Bronnikov, S. B. Fadeev and A. V. Michtchenko, Gen. Rel.
Grav. \textbf{35}, 505 (2003).
\end{thebibliography}
\end{document}